\documentclass[journal]{IEEEtran}

\usepackage{epsf}\epsfverbosetrue
\usepackage{alltt}
\usepackage{subfigure}
\usepackage{graphicx}
\usepackage{picins}
\usepackage{float}
\usepackage{color}
\usepackage{url}
\usepackage{algorithm}
\usepackage{algorithmic}
\usepackage{multirow}
\usepackage{amsfonts}

\centerfigcaptionstrue
\begin{document}

\title{SIMPS: Using Sociology for Personal Mobility}
\author{Vincent~Borrel, Franck~Legendre, Marcelo~Dias~de~Amorim, and Serge~Fdida
\thanks{The authors are with the LIP6/CNRS Laboratory of the Universit{\'{e}} Pierre et Marie Curie~--
Paris VI. Emails: \{borrel,legendre,amorim,sf\}@rp.lip6.fr. This
work has been partially supported by the WIP European project under
contract 27402 and RNRT Airnet.}}

\thispagestyle{plain}

\newtheorem{proposition1}{Definition}

\maketitle

\begin{abstract}
Assessing mobility in a thorough fashion is a crucial step toward
more efficient mobile network design. Recent research on mobility
has focused on two main points: analyzing models and studying their
impact on data transport. These works investigate the
\emph{consequences} of mobility. In this paper, instead, we focus on
the \emph{causes} of mobility. Starting from established research in
sociology, we propose SIMPS, a mobility model of human crowd motion.
This model defines two complimentary behaviors, namely socialize and
isolate, that regulate an individual with regard to her/his own
sociability level. SIMPS leads to results that agree with scaling
laws observed both in small-scale and large-scale human motion.
Although our model defines only two simple individual behaviors, we
observe many emerging collective behaviors (group
formation/splitting, path formation, and evolution). To our
knowledge, SIMPS is the first model in the networking community that
tackles the roots governing mobility.
\end{abstract}


\section{Introduction}

\PARstart{M}{obility} modeling aims at describing in the most
accurate and simplest way the motion of mobile entities. They are
fundamental tools in a large variety of domains, such as physics,
biology, sociology, networking, electronic gaming, and many others.

As of now, there is a growing number of mobility models used in the
design and analysis of communication systems, but how many of them
fully represent the aspects characterizing the mobility of human
beings? This is a fundamental issue, since in many situations the
mobility of communicating and sensing equipments is the reflex of
human mobility. In this paper, we specifically address this
question.

Mobility modeling refers in general to the Random Waypoint model
(RWP), which is the {\it de-facto} standard for both theoretical
analysis and simulation studies.%~\cite{}.
RWP belongs to the same
class as Brownian motion, also called random walk, and has the main
advantages of being simple and analytically tractable. Nevertheless,
the simplicity provided by RWP fails in capturing realistic
behaviors observed in human mobility, as shown by a number of recent
studies~\cite{Perfect05LeBoudec,Jardosh05Real,Johnson04Vehicular,Bettstetter01Smooth}. Fortunately, great advances have
been recently achieved toward more realistic mobility models since
the networking community has decided to investigate mobility in a
finer level of details. A first set of models is based on
expectations of how mobility is performed in particular situations
such as first proposals of campus~\cite{Jardosh05Real} and vehicular
mobility models~\cite{Johnson04Vehicular,Choffnes05Vehicular}.
Another set of proposals tweak RWP parameters with specific
distributions in order to yield more realistic
results~\cite{Jardosh05Real,Bettstetter01Smooth}.

Recent mobility measurements performed both indoor and
outdoor~\cite{chaintreau.infocom06,Kotz02Analysis} enabled the
proposal of trace-based models calibrated with empirical
data~\cite{McNett05Access,tuduce.infocom05,Musolesi04Social}.
Furthermore, a number of analyses show that both contact and
inter-contact
distributions~\cite{Kotz02Analysis,chaintreau.infocom06}, as well as
location popularity distribution~\cite{tuduce.infocom05}, follow
power-law distributions. They also allowed revisiting the realness
of existing models. For example, measurements have confirmed the
presumption that RWP is unable to realistically model human
mobility, since it leads to exponential distributions for both
contact and inter-contact times. Another impact of measurement-based
studies is that it is now possible to reassess mobility model
assumptions. For example, in the valuable work done by Grossglauser
and Tse~\cite{grossglauser.ton02}, the authors assumed i.i.d. random
placement of nodes; this is to be compared to the location popularity
distribution found by Tuduce and Gross~\cite{tuduce.infocom05}.

Despite the increasing number of works questioning the effective
role of mobility, two main issues remain unanswered:

\begin{itemize}

\item {\it Lack of explanation of the process
governing mobility.} Should RWP be used to represent a worst-case
scenario or the uncorrelated displacement of individuals using
different transport facilities? In order to represent more specific
scenarios, a number of models have simply embedded realistic and
higher level features and rules to
RWP~\cite{Jardosh05Real,kn:Zhou99GroupSwarmMobility}.
Yet, neither advanced evidence that they captured realistic displacements.

\item {\it Retained modeling methodology.} Recent proposed
models~\cite{tuduce.infocom05,hsu.mc2r05} have been designed to
artificially match a very limited set of empirical observations. No
clear methodology is applied to evaluate the proposed mobility
models.

\end{itemize}

We argue that a far deeper investigation of the roots governing
mobility is necessary toward realistic mobility modeling. Instead of
simply replaying observed mobility patterns, we propose to rely on
well established theories that tackle the natural process which govern
mobility at its roots. The consequence is the natural emergence of
mobility characteristics found in measurements; this is contrary to
current approaches where these characteristics are artificially
generated. To this end, we revisit the way human mobility modeling
is done by tackling its causes and no more trying to match its
consequences.

In this paper, we propose SIMPS (Sociological Interaction Mobility
for Population Simulation), a mobility model that explores recent
sociological findings driving human interactions: (a) each human has
specific socialization needs, quantified by a target social
interaction level, which corresponds to her/his personal status
({\it e.g.}, age and social class~\cite{ref:StudyVillage,
ref:StudySports}); (b) humans make acquaintances in order to meet
their social interaction
needs~\cite{ref:SocialGraphBuildBehavior,ref:Snijders}. In this
paper, we show that these two components can be translated into a
coherent set of behaviors driving the dynamics of simulated
entities.

For the calibration and validation of the model, we compare the
simulation results with empirical observations obtained from
measurements referenced previously. To our knowledge, this is the
first time a mobility model exhibits such accurate matching with
empirical observations. This is the basis for a high confidence in
the validity of the model.

The remainder of this paper is organized as follows. In
Section~\ref{sec:simps-generalities}, we present some
background required for the definition of our model.
In Section~\ref{sec:simps-details}, we detail the SIMPS model and its parameters.
Then, in Section~\ref{sec:parameters_eval}, we present an extensive analysis of SIMPS parameters.
In Section~\ref{sec:methodology}, we describe the methodology used for the tests performed and discussed in Section~\ref{sec:results}.
In perspective of this analysis, further points are discussed in Section~\ref{sec:furthernotes}.
Finally, in Section~\ref{sec:conc} we conclude this work.

\section{Rationale}
\label{sec:simps-generalities}

In the following, we give the required background to understand our
proposal. We first start by describing the modeling approaches we
have retained. We then reconsider the importance of collective
motion in mobility. Eventually, we describe the sociological basis
upon which our approach relies upon.

SIMPS adopts a mobility modeling approach centered on behavioral
rules. Behavioral mobility models rely on continuously interacting
rules that express atomic behaviors governing mobility. Such an
approach finds great success in other domains such as
physics%~\cite{}
and artificial intelligence.%~\cite{}.
SIMPS defines
two behavioral rules, namely socialize and isolate. These rules
express recent sociological findings driving human interactions. Of
course, no model can realistically integrate all potential behaviors
that drive human motion. In fact, human beings are driven by many
interacting influences, needs and motives driven by schedules,
social ties, to cite a few. Hence, our goal is to {\it(i)} rely on
realistic sociological assumptions {\it (ii)} with a reduced set of
behaviors (simple model as possible) {\it (iii)} still exhibiting
recent distributions observed empirically.

A crucial point in modeling human mobility is to characterize
collective behaviors. The current approach to group modeling does
not go further than proposing correlated motion as in
RPGM~\cite{hong99group} and leaves open the processes behind group
composition (merges) and group splits. SIMPS responds to these
limitations by having emerging collective behaviors results of the
social interactions driven by our two rules. The difficulty is now
to find `realistic' assumptions of social interactions.

\begin{figure}[t]
 \begin{center}
  \epsfxsize=8cm
  \leavevmode\epsfbox{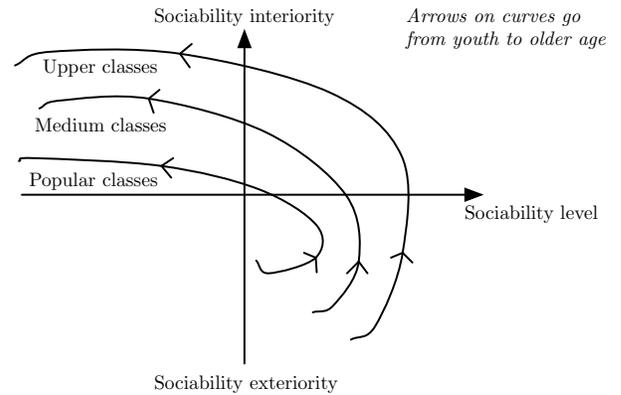}
  \caption{Evolution of the sociability of humans. As can be seen, the characteristics of one's sociability are dependent on intrinsic personal characteristics, such as social class and age, instead of, e.g.,  the situation the person is in.}
  \label{fig:SociabilityEvolution}
 \end{center}
\end{figure}

We base our proposal on the following findings. The first finding,
{\it intrinsicality}, has been expressed concurrently in the
literature by several research
papers~\cite{ref:StudyVillage,ref:StudySports}. It states that the
sociability level of a given person is intrinsic to each person, and
strongly dependent on internal factors ({\it e.g.}, social class and
age). This means that each individual has its own and constant
sociability level at a given period in life that does not depend on
its place in a social network~-- contrary to what one could
assume~-- as shown by Fig.~\ref{fig:SociabilityEvolution}. This
socialization level is translated to a need for social interactions.
The second finding, {\it interactivity}, is derived from a
socialization behavior defined
in~\cite{ref:SocialGraphBuildBehavior,ref:Snijders}, that assesses
that individuals' aim to fulfill their sociological interaction
needs. This can be expressed by building new social ties until the
needs are fulfilled or by satisfying these needs by encountering
already known acquaintances. These acquaintances are defined by
individuals to whom an individual is tightened in a social network.

Yet, social networks have already been used to design mobility
models such as the ones proposed by Musolesi {\em et
al.}~\cite{Musolesi04Social,Musolesi06Community}. In these authors' proposal,
individuals are gathered in clusters by a heuristic which processes
a graph representing social ties. Each cluster is then
affected to a specific region in space. Mobility is generated by
individuals moving from region to region according to a preferential
attachment process.\footnote{In a graph, this process specifies that
incoming nodes create links with already present ones, with a
probability proportional to the latter's degree. While extremely
simple, this process generates graphs with a scale-free node-degree
distribution.} In a companion prior paper, we have already defined
a mobility model based on this process~\cite{Borrel04Attractors}.
Yet, as we will see in the next section, Musolesi {\em et al.}'s
proposal and our model do not exploit the same social basis and
hence differ in their expression and results.

\section{SIMPS: An interaction based mobility model}
\label{sec:simps-details}

In this section, we present SIMPS in detail.

\subsection{Overview}
\label{sec:overview}

SIMPS is a model of the social component of human motion. At the
scale of a simulation, we assume: {\it (i)} fixed social interaction
need per individual and {\it (ii)} fixed social graph representing
social ties between individuals. Hence, the need of social
interactions is satisfied by either encountering acquaintances or escaping
from non-acquaintances. This requires that individuals meet through
spatial displacements (mobility).

The interactions with acquaintances and non-acquaintances can be
translated into a behavioral model. From these behaviors, defined by
rules, individuals join and leave acquaintances. In SIMPS, each
individual is associated with a personal sociability level, which is
the equilibrium point of the social interaction volume she/he tries
to achieve all the time. Each individual is also associated with a
context-aware indicator, namely perceived surround, which indicates
the individual's perception of her/his current socialization volume;
basically, this value depends on the number of surrounding
individuals.

In order to meet the desired sociability level, individuals can
resort to two complementary behaviors: {\it socialize}~-- movements
toward acquaintances~-- and {\it isolate}~-- to escape from
undesired presences. The acquaintances of an individual are
determined by the social graph in which acquaintances are
represented by directed edges. The effects of socialize and isolate
behaviors are, respectively, to raise and lower one's perceived
surround. To activate one of these behaviors, a feedback decision
process estimates, continually and for every individual, her/his
current socialization volume, and compares it to the individual's
own needs.

SIMPS is composed of two parts: {\it social motion influence} and
{\it motion execution unit}. The social motion influence updates an
individual's current behavior to either socialize or isolate. The
{\it motion execution unit} is responsible for translating the
behavior adopted by an individual into motion. We detail these
processes in the following.

\subsection{Social motion influence}
\label{sec:Socialmotioninfluence}

SIMPS simulates the dynamic properties of a population ${\mathcal
P}$ containing $N$ individuals in a two-dimensional plane (although
its expressions can be easily extended to more dimensions). Time
$\tau$ is assumed to be discrete, with steps of $\Delta_{\tau}$.

Individual $i \in {\mathcal P}$ tends to socialize at her/his own
volume, plus or minus a certain variation. This defines the
following two random variables:

\begin{itemize}

  \item $s_{i}$, or the {\it sociability level} of node $i$, is the
  number of individuals that node $i$ aims at being surrounded by.

  \item $t_{i}$, or the {\it tolerance level} of node $i$, is the
  fractional variation of the sociability level under which the
  individual still feels comfortable.

\end{itemize}

Following these random variables, we can define $i$'s {\it social
comfort range}:

\begin{equation}
z_{i} = [s_i(1-t_i), s_i(1+t_i)].
\end{equation}

\begin{figure}[t]
 \begin{center}
  \epsfxsize=8cm]{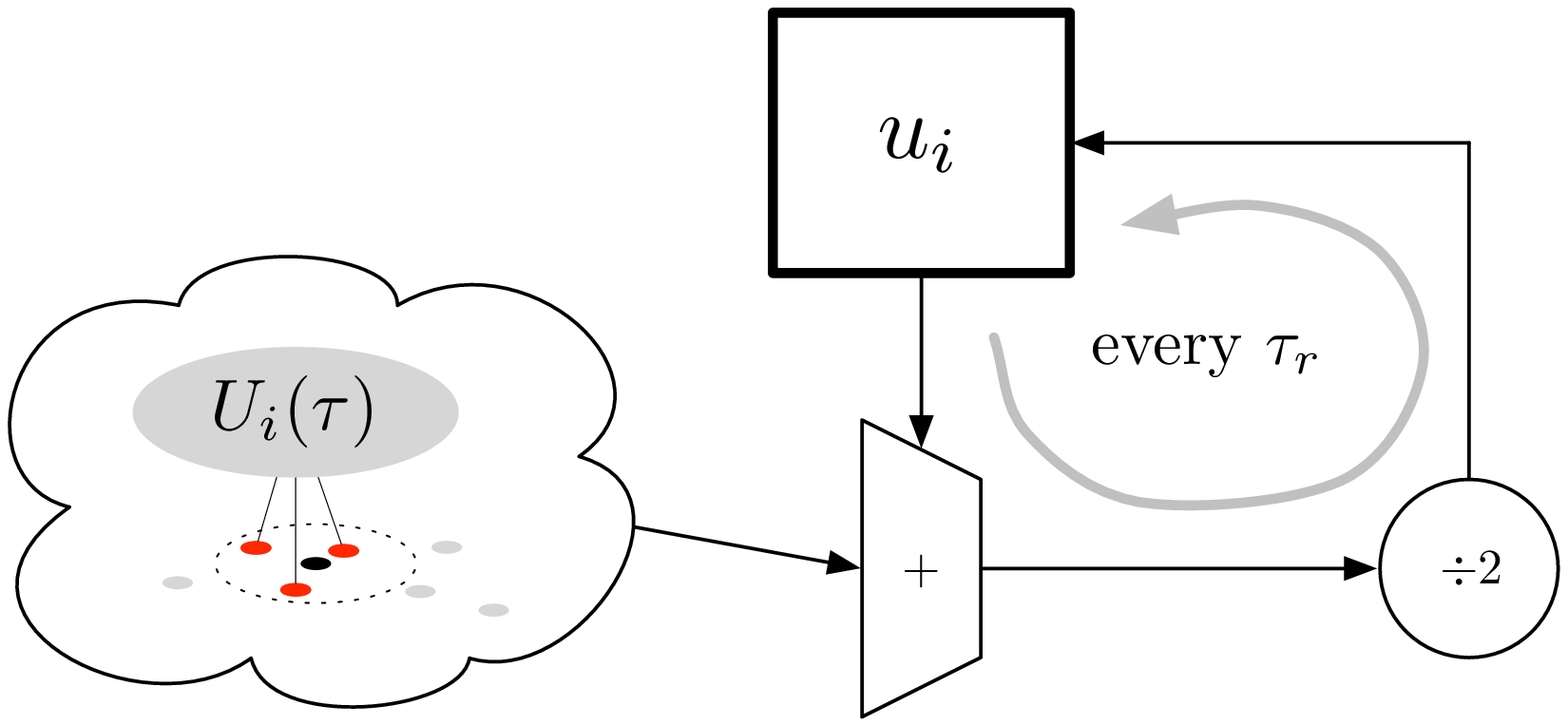}
  \caption{Pseudo-control loop for the computation of the perceived surround.}
  \label{fig:controlloop}
 \end{center}
\end{figure}

According to the theory of
proxemics~\cite{ref:Hall66HiddenDimension}, the social awareness of
an individual is situated in a sphere around her/him, whose radius
$R_{soc}$, namely \emph{social distance}, is approximately 3.5
meters, or 12 feet ({\it cf}., Section~\ref{sec:parameters_eval}).
In this sphere, the perception of nearby individuals is not
immediate, {\it i.e.}, individuals progressively notice the presence
of others. Such a fuzzy perception is called the {\it perceived
surround}, noted $u_{i}$. In order to reproduce this perception, in
SIMPS, individual $i$'s perception is rendered by a pseudo-control
loop as shown in Fig.~\ref{fig:controlloop}. In this loop, the
perceived surround $u_{i}$ is periodically mixed with $U_{i}(\tau)$,
which gives the number of individuals within $i$'s social sphere at
time $\tau$. The period of the pseudo-control loop is called {\it
half-perception time} and noted $\tau_r$.

We can now give expressions that determine the values of the
different variables. Individual $i$'s surround is updated every
$\tau_{r}$ seconds as follows:

\begin{equation}
\label{eq:surround}
 u_{i} = \frac{U_{i}(\tau) + u_{i}}{2},
\label{equ:pseudocontrolloop}
\end{equation}

\noindent where $U_{i}(\tau)$ is given by:

\begin{equation}
\label{eq:u_i}
  U_{i}(\tau) =
  \sum_{
  \begin{array}{c}
    {\scriptstyle j=1}\\
    {\scriptstyle j \neq i}
  \end{array}}^{n}
  P_{i,j}
  \textrm{, where }
  P_{i,j} =
  \left\{
  \begin{tabular}{l}
   $1$, if $|\overrightarrow{ij}| \leq R_{soc},$\\
   $0$, otherwise.
  \end{tabular}
  \right.
 \label{equ:individualscount}
\end{equation}

\noindent In Eq.~\ref{equ:individualscount}, $|\overrightarrow{ij}|$
denotes the Euclidian distance between nodes $i$ and $j$.

\begin{figure}
 \begin{center}
  \epsfxsize=7.5cm
  \leavevmode\epsfbox{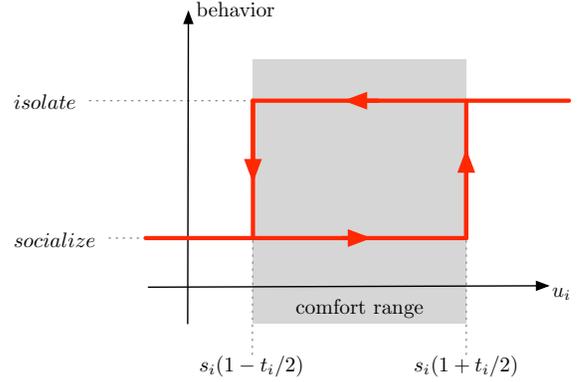}
  \caption{Sharp hysteresis curve for behavior decision process.}
  \label{fig:hysteresis}
 \end{center}
\end{figure}

The perceived surround $u_{i}$ serves as input to the feedback
decision process, which updates the behavior of the individual
according to a sharp hysteresis as shown in
Fig.~\ref{fig:hysteresis}. This hysteresis depends on both the
individual's sociability $s_{i}$ and tolerance level $t_{i}$.

\begin{figure*}[t!]
 \begin{center}
   \epsfxsize=14cm]{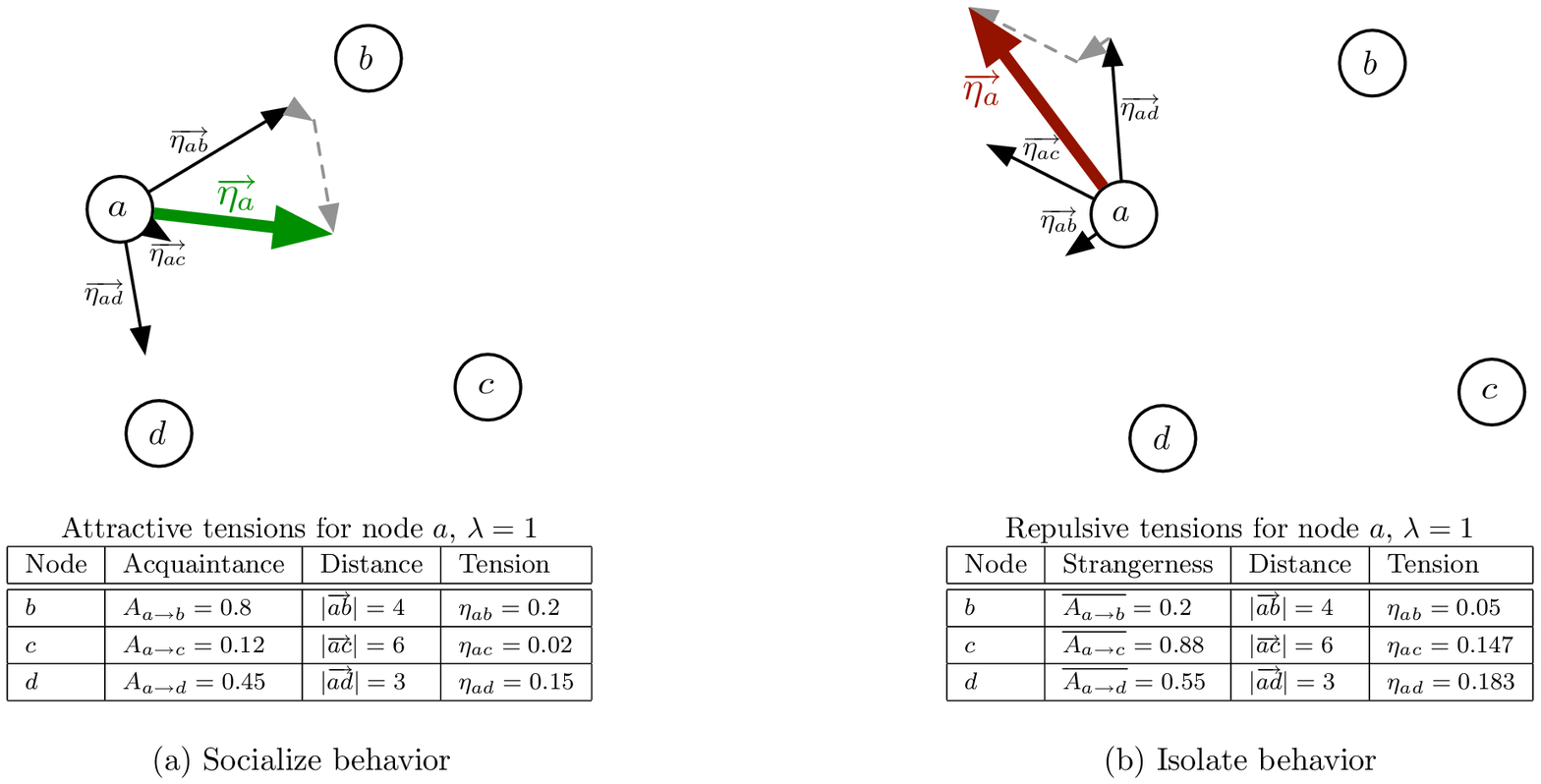}
%   \subfigure[Socialize behavior.]
%    {
%      \label{fig:tensionsSocialize}
%      \epsfxsize=6.0cm
%      \leavevmode\epsfbox{./Figures/TensionsSocialize.eps}
%    }\hspace{1cm}
%    \subfigure[Isolate behavior.]
%    {
%      \label{fig:tensionsIsolate}
%      \epsfxsize=6.0cm
%      \leavevmode\epsfbox{./Figures/TensionsIsolate.eps}
%    }
    \caption{Illustration of the tension calculation process when the individual presents either socialize or isolate behaviors.}
    \label{fig:tensions}
  \end{center}
\end{figure*}

\subsection{The twin social behaviors}
\label{sec:Twinsocialbehaviors}

SIMPS also relies on social graphs from which motion influence
behaviors are derived. Social graphs do not represent physical
proximity, but only relationships among individuals. Nevertheless,
the former influences the latter, since close acquaintances tend to
get physically closer.

In SIMPS, a social graph $G=(V,E)$ is oriented and non-Eulerian.
Vertices represent the nodes in the topology. Links, valuated in the
range $[0;1]$, represent the acquaintances felt from its origin node
toward its destination node: zero means no acquaintance at all ({\it
i.e.}, the destination is stranger to the origin) while one means
high acquaintance ({\it i.e.}, the destination has the maximum
social proximity with the origin).\footnote{Observe that the social
graph does not have to be complete.}

The {\it acquaintance} felt by $i$ toward $j$ is expressed as:
\begin{equation}
A_{i \to j} = \left\{
  \begin{tabular}{ll}
    $f(l_{i \to j})$ & if $l_{i \to j} \in E$ \\
    $0$ & otherwise,\\
  \end{tabular}
\right.
\end{equation}
\noindent where $f(l)$ is the weight of edge $l_{i \to j}$.

Similarly, the {\it strangeness} $\overline{A_{i \to j}}$ felt by
$i$ toward $j$ is defined as:
\begin{equation}
\overline{A_{i \to j}} = 1 - A_{i \to j}.
\end{equation}

We can now precisely define the {\it twin} behaviors of nodes:

\vspace{3mm}
\begin{proposition1}
\label{theo:definition1} ({\bf Socialize}) Individuals are attracted
by acquaintances. The attractive tension
$\overrightarrow{{\eta_a}_{ij}}$ felt by $i$ toward $j$ is a vector
collinear to $\overrightarrow{ij}$, whose magnitude is proportional
to the acquaintance $A_{i \to j}$ and inversely proportional to a
power $\lambda$ of the distance $|\overrightarrow{ij}|$:\footnote{In
our notation, $\hat{\overrightarrow{ij}}$, also expressed as
$\left\|\overrightarrow{ij}\right\|$, is the norm of
$\overrightarrow{ij}$.}

\begin{equation}
\label{equ:attractivetension}
   \overrightarrow{{\eta_a}_{ij}} = \frac{\hat{\overrightarrow{ij}}\, A_{i \rightarrow
   j}}{|\overrightarrow{ij}|^{\lambda}}\cdot
\end{equation}

\end{proposition1}

\vspace{3mm}
\begin{proposition1}
\label{theo:definition2} ({\bf Isolate}) Individuals are repulsed by
strangers. The repulsive tension $\overrightarrow{{\eta_r}_{ij}}$ felt
by individual $i$ toward any other individual $j$ is a vector
collinear to $\overrightarrow{ji}$, whose amplitude is proportional
to the strangeness $\overline{A_{i \to j}}$ and inversely
proportional to a power $\lambda$ of the distance
$|\overrightarrow{ij}|$:%\footnote{Of course, since SIMPS runs over
%Euclidian spaces, $|\overrightarrow{ij}| = |\overrightarrow{ji}|$.}

\begin{equation}
\label{equ:repulsivetension}
    \overrightarrow{{\eta_r}_{ij}} = \frac{\hat{\overrightarrow{ji}}\, \overline{A_{i \rightarrow j}}}{|\overrightarrow{ij}|^{\lambda}}\cdot
\end{equation}

\end{proposition1}

\vspace{5mm} \noindent The $\lambda$ parameter is called {\it
distance fading exponent}.

\vspace{3mm}\noindent {\bf Social motion influence.} Nodes are
either in socialize or isolate mode (as described in
Section~\ref{sec:Socialmotioninfluence}). When in socialize mode, we
have that $\overrightarrow{\eta_{ij}} =
\overrightarrow{{\eta_a}_{ij}}$. On the other hand, when in isolate
mode, node $i$ applies $\overrightarrow{\eta_{ij}} =
\overrightarrow{{\eta_r}_{ij}}$. The vectorial sum
$\overrightarrow{\eta_i} = \sum_{\scriptstyle i \neq j}
\overrightarrow{\eta_{ij}}$ of all attractive or repulsive tensions
give the direction of the willingness, $\overrightarrow{w_i}$, of
$i$'s {\it social motion influence} (as depicted in
Fig.~\ref{fig:tensions}.a and~\ref{fig:tensions}.b):
\begin{equation}
  \label{equ:socialmotioninfluence}
  \overrightarrow{w_{i,\tau}} =
  \left\{
  \begin{tabular}{ll}
    $\hat{\overrightarrow{\eta_i}} e_{i,\tau}$, & if $\left|\overrightarrow{\eta_i}\right| \neq 0$,\\
    $\overrightarrow{0}$, & otherwise,
  \end{tabular}
  \right.
\end{equation}

\noindent where $e_{i,\tau}$, called $i$'s {\it excitation}, is given by:
\begin{equation}
  \label{equ:excitation}
  e_{i,\tau} = \max\left( \left| \frac{u_{i} -
s_i}{s_i t_i} \right|,1 \right).
\end{equation}
The excitation is null when $i$ feels surrounded at his exact
sociability need $s_i$, and progressively increases until attaining
$1$, when $i$'s surround $u_{i}$ falls outside $i$'s comfort range
$z_{i} = [s_i(1-t_i), s_i(1+t_i)]$.

A special instance of Eq.~\ref{equ:socialmotioninfluence} happens
when the sum of tensions is a null vector. Typically, this refers a
situation where an individual is attracted toward many directions at
the same time with a neutral result. In this situation, the
individual hesitates and does not move. This situation, which is
quite rare, is a case of instable equilibrium. The fact that
neighboring individuals move makes this situation very short-lived.

\subsection{Motion execution unit}

The social motion influence is not the sole parameter to have an
impact on human motion. The role of the {\it motion execution unit}
is to comply with two basic parameters governing the physical motion
of individuals: velocity and acceleration. A more complete set of
parameters ({\it e.g.}, collision avoidance or terrain diversity)
could be implemented; however, in order to focus on the social
aspect of mobility, we only consider velocity and acceleration.

Velocities and accelerations are distributed according to two random
variables, respectively $V$ and $A$, whose characteristics will be
discussed later on. Individual $i$ is associated with an
acceleration request, $\overrightarrow{{a_r}_{i,\tau}}$, which is
proportional to $i$'s social motion influence:

\begin{equation}
  \label{equ:accelerationrequest}
  \overrightarrow{{a_r}_{i,\tau}} = {a_{\max}}_i
  \overrightarrow{w_{i,\tau}},
\end{equation}

\noindent where ${a_{\max}}_i$ denotes the maximum scalar acceleration
individual $i$ tolerates.

The acceleration request is applied with respect to $i$'s maximum velocity ${v_{\max}}_i$. Given $i$'s velocity vector at time
$\tau-\Delta_{\tau}$, $\overrightarrow{v_{i,\tau-\Delta_{\tau}}}$,
and the direction $\hat{\overrightarrow{{a_r}_{i,\tau}}}$ of the
acceleration request, we compute the maximum acceleration
$a_{lim_{i,\tau}}$ that can be applied for a duration $\Delta_\tau$ without trespassing ${v_{\max}}_i$. Since SIMPS discretizes time
$\tau$ in steps of $\Delta_{\tau}$, we have:

\begin{equation}
 \overrightarrow{v_{i,\tau}} = \overrightarrow{v_{i,\tau-\Delta_\tau}} + \overrightarrow{a_{i,\tau}}\Delta_{\tau}
 \label{equ:timediscretization}
\end{equation}

\noindent and the speed limit condition:

\begin{equation}
 \left| \overrightarrow{v_{i,\tau}} \right| \leq {v_{\max}}_i.
 \label{equ:speedlimit}
\end{equation}

A short glance at Fig.~\ref{fig:limitacceleration} gives us:

\begin{figure}[t]
 \begin{center}
  \epsfxsize=5.5cm
  \leavevmode\epsfbox{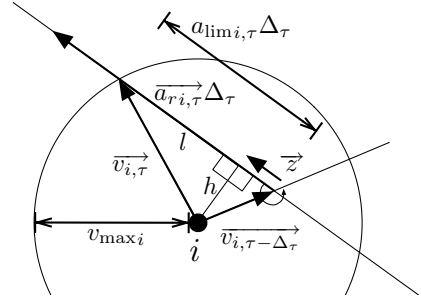}
  \caption{Calculation of ${a_{\lim}}_{i,\tau}$, in relation to $i$'s maximum velocity ${v_{\max}}_i$.}
  \label{fig:limitacceleration}
 \end{center}
\end{figure}

\begin{equation}
  h = \left| \overrightarrow{v_{i,\tau-\Delta_\tau}} \right| \times \left| \sin \left( \widehat{\overrightarrow{{a_r}_{i,\tau}}, \overrightarrow{v_{i,\tau-\Delta_{\tau}}}} \right)
  \right|,
\end{equation}

\begin{eqnarray}
 l & = & \sqrt{{v_{\max}}_i^2-h^2} = \\ \nonumber
   & = & \sqrt{{v_{\max}}_i^2-{\left( \left| \overrightarrow{v_{i,\tau-\Delta_{\tau}}} \right| \times \left| \sin \left( \widehat{\overrightarrow{{a_r}_{i,\tau}}, \overrightarrow{v_{i,\tau-\Delta_{\tau}}}} \right) \right| \right)}^2}
\end{eqnarray}

\noindent and the projection of $\overrightarrow{z}$ on
$\overrightarrow{{a_r}_{i,\tau}}$:

\begin{equation}
 z = \left| \overrightarrow{v_{i,\tau-\Delta_\tau}} \right| \times \left( -\cos \left( \widehat{\overrightarrow{{a_r}_{i,\tau}}, \overrightarrow{v_{i,\tau-\Delta_{\tau}}}} \right)
 \right).
\end{equation}

We have that:

\begin{equation}
   {a_{\lim}}_{i,\tau} \Delta_{\tau} = h + z.
\end{equation}

\noindent We can then obtain:

\begin{eqnarray}
  \label{equ:limitacceleration}
   {a_{\lim}}_{i,\tau} & = & \frac{\sqrt{{v_{\max}}_i^2-{\left( \left| \overrightarrow{v_{i,\tau-\Delta_{\tau}}} \right| \times \left| \sin \left( \widehat{\overrightarrow{{a_r}_{i,\tau}}, \overrightarrow{v_{i,\tau-\Delta_{\tau}}}} \right) \right| \right)}^2}}{\Delta_\tau}\nonumber\\
   & & - \frac{\left| \overrightarrow{v_{i,\tau-\Delta_\tau}} \right| \times \cos \left( \widehat{\overrightarrow{{a_r}_{i,\tau}}, \overrightarrow{v_{i,\tau-\Delta_{\tau}}}} \right)}{\Delta_{\tau}}
\end{eqnarray}

The acceleration request is then satisfied at best by the final
acceleration $\overrightarrow{a_{i,\tau}}$:
\begin{equation}
  \label{equ:finalacceleration}
  \overrightarrow{a_{i,\tau}} = \hat{\overrightarrow{{a_r}_{i,\tau}}} \times \min\left( \left| \overrightarrow{{a_r}_{i,\tau}} \right|, {a_{\lim}}_{i,\tau}
  \right),
\end{equation}
\noindent which is used to update $i$'s current velocity
$\overrightarrow{v_{i,\tau-\Delta_{\tau}}}$ and position
$\overrightarrow{p_{i,\tau}}$:
\begin{equation}
  \label{equ:updatevelpos}
  \left\{
  \begin{tabular}{l}
      $\overrightarrow{v_{i,\tau}} = \overrightarrow{v_{i,\tau-\Delta_{\tau}}} +
      \overrightarrow{a_{i,\tau}} \Delta_{\tau}$, \\
      $\overrightarrow{p_{i,\tau}} = \overrightarrow{p_{i,\tau-\Delta_{\tau}}} + \overrightarrow{v_{i,\tau}} \Delta_{\tau}$.
  \end{tabular}
  \right.
\end{equation}\vspace{2mm}

\section{Evaluation}
\label{sec:parameters_eval}

Table~\ref{table:parameterset} summarizes the parameters that are
internal to the SIMPS mobility generator. They are described in the
following.

\begin{table*}
    \caption{Parameters used by SIMPS.}
    \label{table:parameterset}
    \centering
    \begin{tabular}{|c|l|l|l|l|c|}
       \hline
         \small{Name}
       & \small{Relates to}
       & \small{Type}
       & \small{Description}
       & \small{Value}
       & \small{Investigated} \\
       \hline
       \hline
       \footnotesize{$V$}
       & \footnotesize{Pedestrian motion}
       & \footnotesize{Random variable}
       & \footnotesize{Maximal speed of individuals}
       & \footnotesize{$N(\mu=1.34, \delta^2=0.26)$}
       & \footnotesize{No} \\
       \hline
       \footnotesize{$A$}
       & \footnotesize{Pedestrian motion}
       & \footnotesize{Random variable}
       & \footnotesize{Maximum acceleration of individuals}
       & \footnotesize{$N(\mu=1.3, \delta=0.4)$}
       & \footnotesize{No} \\
       \hline
       \footnotesize{$S$}
       & \footnotesize{Sociability}
       & \footnotesize{Random variable}
       & \footnotesize{Sociability of individuals}
       & \footnotesize{$N(\mu=2.5, \delta=1)$}
       & \footnotesize{Yes} \\
       \hline
       \footnotesize{$T$}
       & \footnotesize{Sociability}
       & \footnotesize{Random variable}
       & \footnotesize{Tolerance of individuals}
       & \footnotesize{Uniform in $[0.1;0.7]$}
       & \footnotesize{No} \\
       \hline
       \footnotesize{$N$}
       & \footnotesize{Social graph}
       & \footnotesize{Integer variable}
       & \footnotesize{Number of nodes/individuals}
       & \footnotesize{$\in \mathbb{N}$}
       & \footnotesize{Yes} \\
       \hline
        \footnotesize{$D$}
       & \footnotesize{Social graph}
       & \footnotesize{Real variable}
       & \footnotesize{Average node outdegree}
       & \footnotesize{$\in [0;N-1]$}
       & \footnotesize{Yes} \\
       \hline
        \footnotesize{{\it Graph type}}
       & \footnotesize{Social graph}
       & \footnotesize{Enumerated value}
       & \footnotesize{Graph type}
       & \footnotesize{in \{Natural,Random,Scale-Free\}}
       & \footnotesize{Yes} \\
       \hline
        \footnotesize{$R_{soc}$}
       & \footnotesize{Human perception}
       & \footnotesize{Real variable}
       & \footnotesize{Social radius}
       & \footnotesize{$3.5$ meters}
       & \footnotesize{Yes} \\
       \hline
        \footnotesize{$\tau_r$}
       & \footnotesize{Human perception}
       & \footnotesize{Real variable}
       & \footnotesize{Half-perception time}
       & \footnotesize{$4$ seconds}
       & \footnotesize{Yes} \\
       \hline
        \footnotesize{$\lambda$}
       & \footnotesize{Human perception}
       & \footnotesize{Real variable}
       & \footnotesize{Distance fading exponent}
       & \footnotesize{$\in [0;3]$}
       & \footnotesize{Yes} \\
       \hline
        \footnotesize{{\it Space}}
       & \footnotesize{Space}
       & \footnotesize{Enumerated value}
       & \footnotesize{The space where motion happens}
       & \footnotesize{in \{Infinite, Periodic Square\}}
       & \footnotesize{Yes} \\
       \hline
        \footnotesize{$L$}
       & \footnotesize{Space}
       & \footnotesize{Real variable}
       & \footnotesize{Size of periodic shape}
       & \footnotesize{$\in [0;+\infty[$}
       & \footnotesize{Yes} \\
       \hline
        \footnotesize{$\tau_{\max}$}
       & \footnotesize{Time}
       & \footnotesize{Real variable}
       & \footnotesize{Total time considered}
       & \footnotesize{$\in [0;+\infty[$}
       & \footnotesize{Yes} \\
       \hline
        \footnotesize{$\Delta_{\tau}$}
       & \footnotesize{Time}
       & \footnotesize{Real variable}
       & \footnotesize{Time quantization step}
       & \footnotesize{$\in [0;\tau_{\max}[$}
       & \footnotesize{Yes} \\
       \hline
    \end{tabular}
\end{table*}

\subsection{Pedestrian motion characteristics}
\label{subsec:ped-motion}

This parameter dictates the distributions of the random variables
$V$ and $A$, respectively the velocity and acceleration of the
individuals. According to results published by Henderson
in~\cite{Henderson71Crowd}, velocity $V$ for pedestrians is set to
follow a normal law $N(\mu=1.34, \delta^2=0.26)$. Acceleration
distribution is harder to gauge. Considering that a human can switch
from immobile position to walking in the order of the second, we
empirically set it to follow a similar normal law $N(\mu=1.3,
\delta^2=0.4)$.\footnote{While we do not focus on this aspect here,
some of our tests showed that doubling or halving acceleration
settings do not significantly change the outcome of the mobility.}

\subsection{Social interactions characteristics}
\label{subsec:soc-int}

These characteristics govern the interactions between individuals.
Random variables $S_{i}$ and $T_{i}$, introduced in
Section~\ref{sec:Socialmotioninfluence}, denote, respectively, the
volume of social interaction required by each individual
(sociability) and the variation she/he tolerates on this volume of
interaction (tolerance). To assess $S_{i}$, we look into the group
size distribution found in~\cite{Henderson71Crowd}. The results are
expressed as a Poisson law ($\lambda=2.5$) of discrete values. Since
the sociability distribution in SIMPS is a continuous function, we
translate the discrete Poisson law into a normal law $N(\mu=2.5,
\delta^2=1)$,\footnote{These laws are not exactly equivalent, but
this is the closest form we can find in literature.} for which we
investigate in this paper the effect of variation of its first
moment. Furthermore, the tolerance on this sociability is set to a
uniform distribution in $[0.1;0.7]$ ({\it i.e.}, between $10\%$ and
$70\%$ tolerance on the sociability).

The social graph defines acquaintances between individuals, which
are used to compose the twin behaviors. This graph is parameterized
by the number of nodes, $N$, the average node degree $D$, and the
type, which can be natural (when drawn from real traces) or
synthetic ({\it e.g.}, Erd\"os-R\'enyi random graph, exponential
node-degree distributed graph, or Albert-Barab\'asi scale-free graph
with power-law node degree distribution). Average node degree for
various measured social graphs vary more than one order of magnitude
depending on the subject of the social graph. For example, sexual
contact graphs exhibit a low average node degree ($D \approx 2$),
followed by phone calls graphs ($D \approx 3$), blogs ($D \approx
14$), up to actor collaboration ($D \approx 61$). Since nodes in the
graph represent mobile individuals, the size of the graph gives the
population size to which the random variables $V$, $A$, $S$, and $T$
apply.

\subsection{Human perception characteristics}
\label{subsec:hum-perc}

This parameter defines the way human beings perceive their
environment. They are in number of three: social distance,
half-perception time, and distance fading exponent. Proxemics,
defined by E. T. Hall~\cite{ref:Hall66HiddenDimension}, stipulates
that relations between humans are dependent on the distance
separating them. The social distance is the physical distance under
which social transactions and interactions occur. In the United
States, this distance was measured to be around $12$ft. (or $3.5$m);
however, this value was found to vary from about half to several
times this distance, depending on cultural variations, and also on
spatial constraints, such as typically crowded areas. In SIMPS, this
value is directly translated into $R_{soc}$, the distance used for
one's current socialization estimation. While $3.5$m is the typical
value we set for our tests, we also investigate the impact of its
variations.

The half-perception time $\tau_r$ regulates the pseudo-control loop
described in Section~\ref{sec:Socialmotioninfluence}. It describes
the time an individual takes to perceive changes in her/his
neighborhood. Although, to our knowledge, this parameter has not
been much investigated, some documents in the literature show that
the perception time may range from hundreds of milliseconds to tens
of seconds~\cite{ref:WelfordReactionTimes,ref:TriggsDrivers}. SIMPS is about pedestrian motion, which
is a more relaxed environment; in such a context, users spend more
time to react, and reaction times are considered in the order from
around one second to tens of second.

One of the particularities of physical motion is that each movement
has a cost. In SIMPS, this is taken into account by the distance
fading exponent $\lambda$ ({\it cf.},
Section~\ref{sec:Twinsocialbehaviors}), which defines the cost an
individual associates to the distance that separates her/him from
another individual. Basically, when $\lambda = 1$, it means that the
distance has a first order impact on the result (the cost of a motion is
considered linear to the distance), while $\lambda = 2$ means that
the distance has a second order impact (the cost is considered to be in square of
the distance). In Section~\ref{sec:results}, we will investigate
$\lambda$ in detail.

\subsection{Spatial characteristics}
\label{subsec:spat-charac}

This parameter describes the space in which individuals evolve. The
boundary conditions can be of three types: infinite, finite, and
periodic. If finite, the topology can be a square, a hexahedron, a
disc, a bitmap, or a parametric space given by a twin set of
polygons (presence zone polygons minus obstacle polygons). If
periodic, the topology can be a square (with toroidal boundary
mapping), a set of hexahedrons (with cell-like boundary mapping), or
a pair of discs (with bi-hemispheric boundary mapping). In the
remainder of this paper, we investigate the properties exhibited by
SIMPS alone. Aiming at the simplest scenarios, we will restrict our
study to the infinite and periodic square (toroidal mapping) cases,
in which the influence of square side $L$ will be investigated.

\subsection{Time characteristics}
\label{subsec:time-charac}

Time characteristics concern the total duration $\tau_{\max}$ for
which motion is considered, and the time quantization step
$\Delta_{\tau}$ used for motion rendering. These two values,
although more related to implementation than to model definition,
are of prime concern since their choice can directly influence the
outcome of the synthesized motion. It is then of major importance to
distinguish inherent characteristics of our model from eventual
effects on its outcomes due to  time sampling. In the analysis
below, we explore the effect of time quantization and total considered duration on the results of SIMPS.

\section{Methodology}
\label{sec:methodology}

In this section, we wish to highlight the outcome of our approach,
which comforts recent observations of power-laws. The most common
observation of power-law distributions lies in the
contact/inter-contact durations. It is exhibited by studies
conducted on WiFi-enabled devices at ETH Zurich, Dartmouth, and
UCSD. Recently, a specific set of experiments conducted by Cambridge
University, UK, in collaboration with Intel, traced the contact and
inter-contact duration distributions between mobile users carrying
iMotes Bluetooth devices~\cite{chaintreau.infocom06}. Although at a
lower communication scale than previous studies, the three
experiments conducted in~\cite{chaintreau.infocom06} (all three at
different times and locations, and with different users) showed
strikingly similar observation of the scale-free characteristics of
human contacts. Contact and inter-contact durations have been
observed to follow power laws whose exponents were situated
respectively around $-1.5$ and $-0.6$, with cut-offs related to the
durations of the observations.

In order to investigate social interactions between individuals, we
simulate human mobility with conditions similar to~\cite{chaintreau.infocom06}.
In the iMotes experiments, users
carried Bluetooth-enabled devices, which periodically recorded the
presence of other BlueTooth-enabled devices, such as other iMotes,
PDAs, mobile phones, or laptops. A contact situation between
individuals was asserted as soon as the presence of one node was
felt by the other one, and an inter-contact asserted as soon as two
or more consecutive measures did not show the presence of a
previously seen node. The theoretical range of BlueTooth is around
$10$ meters. We consider however that a more realistic $6$-meter
range is a valid assertion for sensing range in most situations. In
this way, we chose to simulate a simplistic range-based contact
condition based on a maximum distance of $6$ meters separating
individuals.

\begin{table*}
    \caption{Summary of test settings.}
    \label{table:TestSettings}
    \centering
    \begin{tabular}{|l|c|c|c|c|c|c|c|c|c|c|}
       \hline
         \small{Aspect}
       & \small{$\Delta_{\tau}$}
       & \small{$\tau_{max}$}
       & \small{$\left<S\right>$}
       & \small{Graph type}
       & \small{$D$}
       & \small{$\lambda$}
       & \small{$R_{soc}$}
       & \small{Space}
       & \small{$L$}
       & \small{$\tau_r$} \\
       \hline
       \hline
       \hspace*{-1mm}\footnotesize{Social graph  type}
       & \footnotesize{$1$s}
       & \footnotesize{$3600$s}
       & \footnotesize{$2.5$}
       & \hspace*{-1.6mm}\footnotesize{$[$Random,SF$]$}\hspace*{-1.6mm}
       & \footnotesize{$5$}
       & \footnotesize{$1$}
       & \footnotesize{$3.5$m}
       & \footnotesize{Periodic}
       & \footnotesize{$200$m}
       & \footnotesize{$4$s} \\
       \hspace*{-1mm}\footnotesize{Avg. node  degree}
       & \footnotesize{$1$s}
       & \footnotesize{$3600$s}
       & \footnotesize{$2.5$}
       & \footnotesize{Scale-Free}
       & \hspace*{-1.6mm}\footnotesize{$[2,5,15,50]$}\hspace*{-1.6mm}
       & \footnotesize{$1$}
       & \footnotesize{$3.5$m}
       & \footnotesize{Periodic}
       & \footnotesize{$200$m}
       & \footnotesize{$4$s} \\
       \hline
       \hspace*{-1mm}\footnotesize{Sociability}
       & \footnotesize{$1$s}
       & \footnotesize{$3600$s}
       & \hspace*{-1.6mm}\footnotesize{$[1,2.5,10]$}\hspace*{-1.6mm}
       & \footnotesize{Scale-Free}
       & \footnotesize{$5$}
       & \footnotesize{$1$}
       & \footnotesize{$3.5$m}
       & \footnotesize{Periodic}
       & \footnotesize{$200$m}
       & \footnotesize{$4$s} \\
       \hline
       \hspace*{-1mm}\footnotesize{Socialize only}
       & \footnotesize{$1$s}
       & \footnotesize{$3600$s}
       & \footnotesize{$2.5$}
       & \footnotesize{Scale-Free}
       & \footnotesize{$5$}
       & \footnotesize{$1$}
       & \footnotesize{$0$}
       & \footnotesize{Periodic}
       & \footnotesize{$200$m}
       & \footnotesize{$4$s} \\
       \hspace*{-1mm}\footnotesize{Isolate only}
       & \footnotesize{$1$s}
       & \footnotesize{$3600$s}
       & \footnotesize{$0$}
       & \footnotesize{Scale-Free}
       & \footnotesize{$5$}
       & \footnotesize{$1$}
       & \footnotesize{$> \sqrt{2} \times L$}
       & \footnotesize{Periodic}
       & \footnotesize{$200$m}
       & \footnotesize{$4$s} \\
       \hline
       \hspace*{-1mm}\footnotesize{Social distance}
       & \footnotesize{$1$s}
       & \footnotesize{$3600$s}
       & \footnotesize{$2.5$}
       & \footnotesize{Scale-Free}
       & \footnotesize{$5$}
       & \footnotesize{$1$}
       & \hspace*{-1.6mm}\footnotesize{$[1,3.5,15]$m}\hspace*{-1.6mm}
       & \footnotesize{Periodic}
       & \footnotesize{$200$m}
       & \footnotesize{$4$s} \\
       \hspace*{-1mm}\footnotesize{Reaction time}
       & \footnotesize{$1$s}
       & \footnotesize{$3600$s}
       & \footnotesize{$2.5$}
       & \footnotesize{Scale-Free}
       & \footnotesize{$5$}
       & \footnotesize{$1$}
       & \footnotesize{$3.5$m}
       & \footnotesize{Periodic}
       & \footnotesize{$200$m}
       & \hspace*{-1.6mm}\footnotesize{$[1,4,20]$s}\hspace*{-1.6mm} \\
       \hspace*{-1mm}\footnotesize{Distance cost}
       & \footnotesize{$1$s}
       & \footnotesize{$3600$s}
       & \footnotesize{$2.5$}
       & \footnotesize{Scale-Free}
       & \footnotesize{$5$}
       & \hspace*{-1.6mm}\footnotesize{$[0,1,2,3]$}\hspace*{-1.6mm}
       & \footnotesize{$3.5$m}
       & \footnotesize{Periodic}
       & \footnotesize{$200$m}
       & \footnotesize{$4$s} \\
       \hline
       \hspace*{-1mm}\footnotesize{Space: infinite}
       & \footnotesize{$1$s}
       & \footnotesize{$3600$s}
       & \footnotesize{$2.5$}
       & \footnotesize{Scale-Free}
       & \footnotesize{$5$}
       & \footnotesize{$1$}
       & \footnotesize{$3.5$m}
       & \footnotesize{Infinite}
       & \footnotesize{$\infty$}
       & \footnotesize{$4$s} \\
       \hspace*{-1mm}\footnotesize{Space: periodic}
       & \footnotesize{$1$s}
       & \footnotesize{$3600$s}
       & \footnotesize{$2.5$}
       & \footnotesize{Scale-Free}
       & \footnotesize{$5$}
       & \footnotesize{$1$}
       & \footnotesize{$3.5$m}
       & \footnotesize{Periodic}
       & \hspace*{-1.6mm}\footnotesize{$[20,200,2000]$m}\hspace*{-1.6mm}
       & \footnotesize{$4$s} \\
       \hline
       \hspace*{-1mm}\footnotesize{Total duration}
       & \footnotesize{$1$s}
       & \hspace*{-1.6mm}\footnotesize{$[600$s$,1$h$,10$h$]$}\hspace*{-1.6mm}
       & \footnotesize{$2.5$}
       & \footnotesize{Scale-Free}
       & \footnotesize{$5$}
       & \footnotesize{$1$}
       & \footnotesize{$3.5$m}
       & \footnotesize{Periodic}
       & \footnotesize{$200$m}
       & \footnotesize{$4$s} \\
       \hspace*{-1mm}\footnotesize{Time quantization}
       & \hspace*{-1.6mm}\footnotesize{$[0.1,1,10]$s}\hspace*{-1.6mm}
       & \footnotesize{$3600$s}
       & \footnotesize{$2.5$}
       & \footnotesize{Scale-Free}
       & \footnotesize{$5$}
       & \footnotesize{$1$}
       & \footnotesize{$3.5$m}
       & \footnotesize{Periodic}
       & \footnotesize{$200$m}
       & \footnotesize{$4$s} \\
       \hline
    \end{tabular}
\end{table*}

\section{Results and discussion}
\label{sec:results}

Tests on the outcome of SIMPS mobility regarding contact and
inter-contact distributions have been conducted over a population of
$N = 100$ individuals. The general parameters used by SIMPS are
shown in Table~\ref{table:TestSettings} while the different values
of the simulations we conducted are depicted in
Table~\ref{table:parameterset}.

\subsection{Influence of social graph}

\begin{figure}[t]
  \begin{center}
  \hspace*{-2.4mm}
  \epsfxsize=8.9cm
  \leavevmode\epsfbox{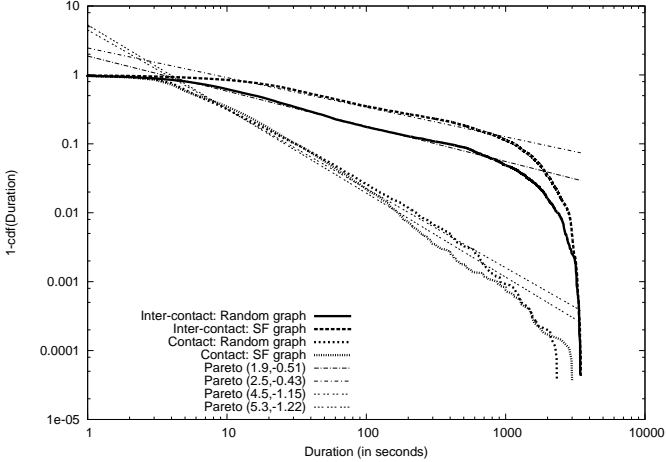}
  \caption{Effect of the underlying graph on the outcome of the SIMPS mobility.
  Contact and inter-contact duration distributions taken with two inputs:
  Albert-Barab\'asi scale-free graph with power law node degree distribution
  and Erd\"os-R\'enyi random graph with exponential node degree distribution.}
  \label{fig:graphinfluence}
  \end{center}
\end{figure}

The first aspect to gauge is the behavior of the model in the
general case, and its dependence with the underlying graph
structures. To this effect, the distributions of contact and
inter-contact durations of Bluetooth-carrying nodes subject to SIMPS
mobility are plotted, for both random and scale-free graphs, in
Fig.~\ref{fig:graphinfluence}. As we can observe, the contact and
inter-contact duration distributions follow power-laws in both
cases, and with very close exponents. The distribution of
inter-contact is a bit below the one of the scale-free graph, while
contact duration distributions are very close for both graph
structures. All these distributions experiment an exponential
cut-off around $500$ seconds. As will be seen later on, these
cut-offs are due to the duration of the simulation.

The similarity of the mobility patterns obtained with scale-free and
exponential graphs is noticeable, and both correspond to the same power law of exponent
$\alpha = 1.2$. Inter-contact distributions are also very similar
for both graph types, with a slightly sharper cut-off for the
scale-free social graph. We conclude from this that the graph
structure has minor influence relatively to the emergence of the
power-law contact and inter-contact distributions. This result is
very important and surprising: {\it although social graph are
notable for their scale-free degree distribution, scale-free contact
and inter-contact distributions  are not due to this feature, and
emerges from the social motion}.

\begin{figure}
  \begin{center}
  \hspace*{-2.4mm}
  \epsfxsize=8.9cm
  \leavevmode\epsfbox{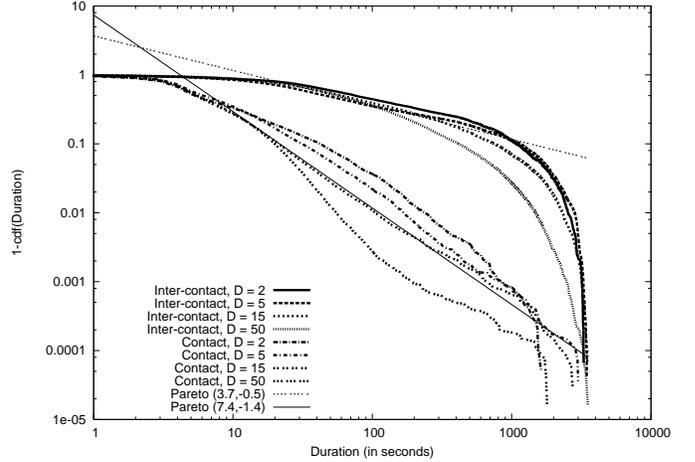}
  \caption{Effect of the variation of average node degree $D$ on the outcome of SIMPS.}
  \label{fig:NodeDegreeVariation}
  \end{center}
\end{figure}

The average node-degree $D$ of the social graph can vary significantly
depending on the scenario. We verify then how SIMPS behaves under
different average node degrees. Tests have been run for four
different values of $D$: $2$, $5$, $15$, and $50$. These values
span a broad range of situations encompassing most social graphs
studied in literature. The results are plotted in
Fig.~\ref{fig:NodeDegreeVariation}. We observe that contact and
inter-contact distributions remain quite stable in their power-law
nature, despite the important changes in the underlying social
graph. The exponents in both distributions tend to increase slightly
in function of $D$.\footnote{Notice the extreme case where $D = 50$
and a node is, in average, an acquaintance of half the whole
population. In this case, the inter-contact distribution tends to a
Weibull form. This tendency can also be found in
UCSD~\cite{McNett05Access} and Dartmouth~\cite{Kotz02Analysis}
studies.}

The two previous observations give us high confidence in the
robustness of SIMPS, and its reasonable independence relatively to
the social graph used. For this reason, we decide to use
Albert-Barab\'asi scale-free social graphs for the remainder of this
paper. These are considered to be closer to real social graphs than
Erd\"os-R\'enyi random graphs. The default $D$ value used, unless
needed otherwise, will be $5$.

\subsection{Influence of sociability}

\begin{figure}[t]
  \begin{center}
  \hspace*{-2.4mm}
  \epsfxsize=8.9cm
  \leavevmode\epsfbox{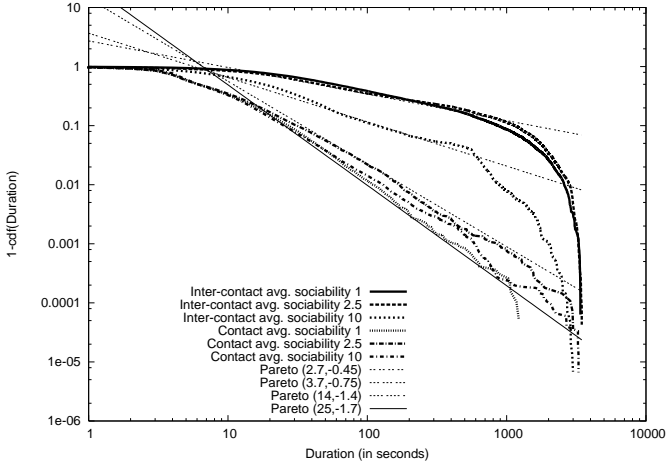}
  \caption{Effect of the average sociability on the outcome of the
  SIMPS mobility, for contact and inter-contact duration
  distributions of mobile nodes carrying Bluetooth-like devices.}
  \label{fig:graphSociability}
  \end{center}
\end{figure}

The next parameter we must verify for a social model is the
variation of the social influence. We performed tests for three
different values of $\left<S\right>$, the first moment of random variable $S$, for which the distributions are shown in Fig.~\ref{fig:graphSociability}. We
can see from the graph that although the slope of both distributions
change with the settings, their nature remains as a power-law with
cut-off. For larger values of $\left<S\right>$, in which humans have
higher needs for socialization, the inter-contact distribution
changes the most. This is an important phenomenon that we will
explore hereafter: the power-law nature of both distributions
outputted by SIMPS seems strictly dependent on the balanced presence
of {\it both} socialize and isolate behaviors.

\subsection{Separate effects of socialize and isolate behaviors}

\begin{figure}
  \begin{center}
  \hspace*{-2.4mm}
  \epsfxsize=8.9cm
  \leavevmode\epsfbox{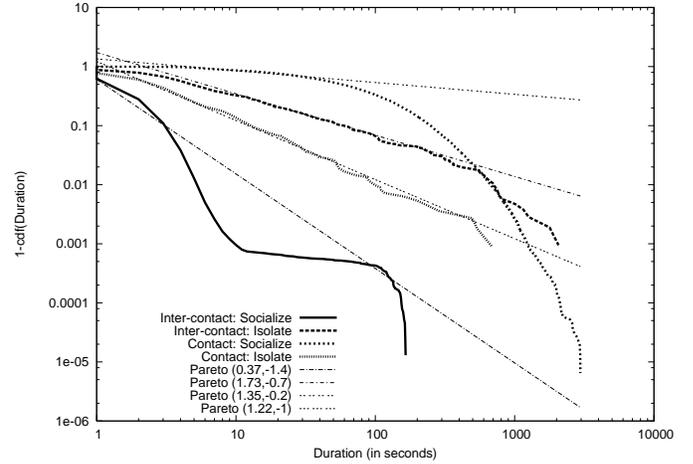}
  \caption{Separate effect of {\it socialize} and {\it isolate} behaviors on SIMPS results.}
  \label{fig:IsolateSocialize}
  \end{center}
\end{figure}

We evaluate now the contribution of each of the twin social
behaviors on mobility. To this end, we set the model for three runs.
The first considers only the socialize behavior, the second
considers only the isolate behavior, and the third considers both
behaviors together. Setting a single behavior is accomplished by
changing $R_{soc}$, the social distance (defined in
Section~\ref{sec:Socialmotioninfluence}). Setting $R_{soc} = 0$
means that individuals always want to socialize. Setting $R_{soc} >
\sqrt{2} \times L$ and distributing $S$ and $T$ so as no individual
tolerates a surround of more than $N-1$ individuals imposes that all
individuals will feel over-socialized and want to isolate. The
corresponding results are shown in Fig.~\ref{fig:IsolateSocialize}.

We can observe from the graph the clear influence of the isolate
behavior on the power-law distribution. Both contact and
inter-contact duration distributions exhibit neat scaling laws. It is however not sufficient to explain the strong exponent difference between these two distributions, characteristic of all observations. The
outcome of the socialize behavior is by far the more interesting~--
inter-contact distribution is far below contact distribution, and
follows two power law sections separated by a sudden decrease. Such
a staircase-like feature could explain the presence of two smaller
bumps barely visible as an outcome of SIMPS, and also present in
various real-life measures, such as in~\cite{chaintreau.infocom06}.
In order to ensure that these bumps were not the result of the
periodic space used, an additional run has been performed in an
infinite space. The results exhibit the same characteristic. Another
interesting emergent behavior is that, although individuals are most
of the time in socialize mode, the global outcome of SIMPS tends
more toward the characteristics exhibited by the isolate behavior.
This seems to indicate that the mixture of both behaviors is very
different from their average, hence pointing the emergence of a
different mechanism from their interplay.

\subsection{Influence of human perception}

\begin{figure}
  \begin{center}
  \hspace*{-2.4mm}
  \epsfxsize=8.9cm
  \leavevmode\epsfbox{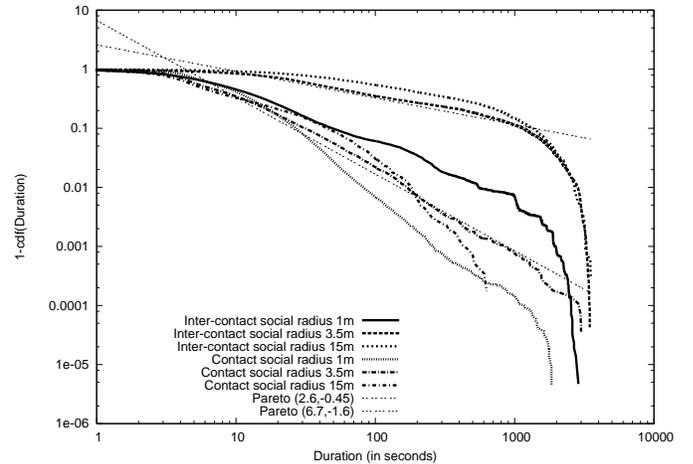}
  \caption{Effect of the social distance on SIMPS outcome.}
  \label{fig:VariationRSoc}
  \end{center}
\end{figure}

\subsubsection{Social distance}

Human perception also has direct influence on mobility. As discussed
earlier in this paper, the typical social distance $R_{soc}$ between
human beings was found to be around $3.5$ meters for U.S.
citizens~\cite{ref:Hall66HiddenDimension}, subject to variations
depending on sociocultural background (\emph{e.g.} smaller in Latin
countries, bigger for rural areas) and context (\emph{e.g.} reduced
in crowded areas, sport events). In order to assess the effect of
perception on mobility, a test has been run with 3 different values
of $R_{soc}$: $1$~m, $3.5$~m, and $15$~m. The results are shown in
Fig.~\ref{fig:VariationRSoc}.

As seen with the case where $R_{soc}$ is reduced to $1$~m, contact
and inter-contact duration distributions still show a power-law
feature with cut-off, but with a higher exponent. When $R_{soc}$ is
raised to $15$~m, both distributions experience a slight decrease in
its exponent; however, the cut-off arrives much sooner in the
contact distribution. This might be due to border effects, when the
size of the social sphere approaches the total space considered for
mobility (as we will see in Section~\ref{subsec:space-variations}).
While interesting for our study, it is unlikely that this extreme
case happens in reality, since proxemics states that the variation
of social distance is related to the context an individual is
immersed into; for instance, social distance is reduced in crowded
areas.

\begin{figure}
  \begin{center}
  \hspace*{-2.4mm}
  \epsfxsize=8.9cm
  \leavevmode\epsfbox{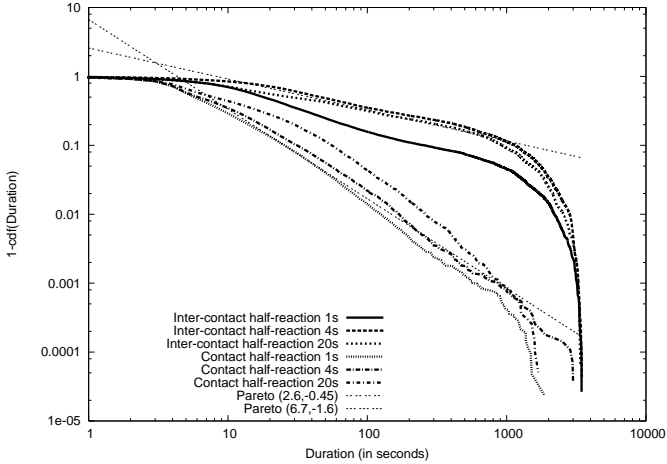}
  \caption{Effect of the half-perception time on SIMPS outcome.}
  \label{fig:VariationTauR}
  \end{center}
\end{figure}

\subsubsection{Half-perception time}

We now investigate the impact of how quick human beings perceive
changes in their immediate surroundings. We performed tests with
three possible value of the half-perception time, $\tau_r$: $1$~s,
$4$~s, and $20$~s. The results are shown in
Fig.~\ref{fig:VariationTauR}.

Although all values of $\tau_r$ lead to close results, we note two
tendencies. On the one hand, when perception time is minimal, the
inter-contact distribution exhibits lower values, but more
frequently. This is in part due to the fact that individuals rapidly
try to escape from crowds and come back to meet acquaintances. A
similar observation can be drawn for the contact distribution, but
with lower impact. On the other hand, when perception time is at the
longest bound, inter-contact distribution does not change that much,
while contact distribution shows a neat decrease in the proportion
of shorter contacts.

It finally appears that the half-perception time of $4$~s (chosen
for the rest of our simulations) does not lead to any of the border
effects observed above. It results in strict power-laws and clear
distinction between contact and inter-contact distributions, which
are closer to the results obtained by Chaintreau {\it et
al.}~\cite{chaintreau.infocom06}.

\begin{figure}
  \begin{center}
  \hspace*{-2.4mm}
  \epsfxsize=8.9cm
  \leavevmode\epsfbox{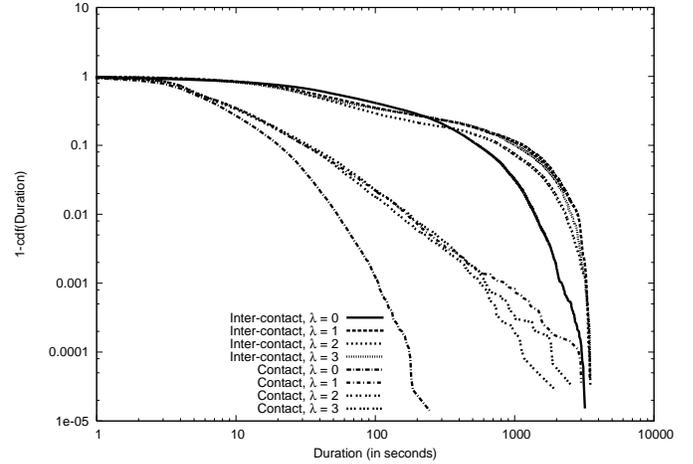}
  \caption{Effect of the distance-associated cost estimation on SIMPS outcome.}
  \label{fig:VariationLambda}
  \end{center}
\end{figure}

\subsubsection{Distance fading exponent $\lambda$}

One of the particularities of physical motion is that each movement
has a cost. In SIMPS, this is taken into account by the distance
fading exponent, $\lambda$, which is used by each individual to
weight the interest of meeting/avoiding another individual
relatively to the distance separating them.

We ran tests with four different values of $\lambda$: $0$ (distance
has no influence on attraction between individuals), $1$ (distance
plays linearly), $2$ (distance plays as a square), and $3$ (distance
plays as a cube).\footnote{Note that $\lambda$ can also take real
values in SIMPS expression.} The results are shown in
Fig.~\ref{fig:VariationLambda}. The main outcome of these tests is
that when distance is not taken into account ($\lambda = 0$),
mobility is mainly characterized by a Weibull distribution, instead
of a power-lay. While still showing scale-invariance in time,
Weibull distributions are less characteristic of empirical traces,
which tend more to a power-law.

Furthermore, the order at which distance is used in cost estimation
does not fundamentally change the outcomes of SIMPS. The results
indicate that distance plays a direct role in human displacement
decisions, confirming results shown in~\cite{Borrel04Attractors}.

\subsection{Influence of the simulation space}
\label{subsec:space-variations}

\begin{figure}
  \begin{center}
  \hspace*{-2.4mm}
  \epsfxsize=8.9cm
  \leavevmode\epsfbox{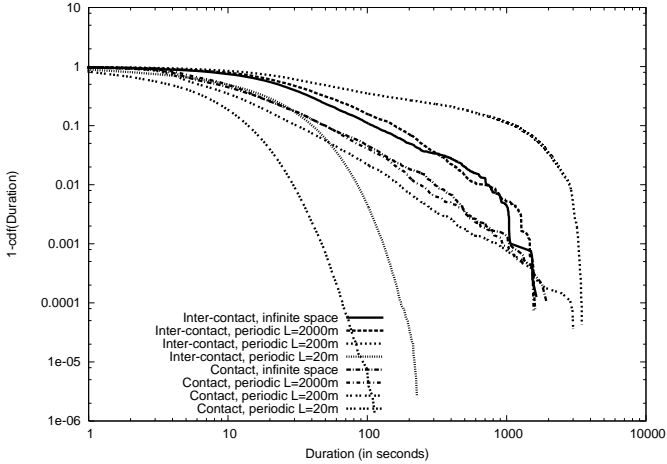}
  \caption{Effect of the considered space on SIMPS results.}
  \label{fig:Spacevariation}
  \end{center}
\end{figure}

Individuals present different mobility characteristics depending on
the space they evolve in. We explore here two parameters defining
motion space. The first one, namely space type, defines if
individual evolve in free, limited, or periodic space. In the case
of limited space, two options can be considered to manage the
situation where individuals attain a border: (a) nodes stop when
they face a border (obstruction); (b) nodes' directions are
symmetrically reflected against limits (reflection). In periodic
spaces, these latter are considered as a torus~-- nodes escaping one
border are replaced in the opposite border of the space. This model
has the advantage of leading to regular average spatial density.

Contrary to the experiments described earlier in this paper, which
ran over periodic spaces, we consider now more realistic free
spaces. The goal is to investigate if our observations still hold.
In addition, in the case of a non-infinite space, we also consider
the impact of the density on the results, since it directly
influences the equilibrium between the quantities of socializing and
isolating individuals. We vary density with the inverse square of
side $L$. The first run is performed in free-space, with users
initially uniformly disposed in a square of 200~m on side, while the
other simulations are performed in a periodic square space with
sides varying between $20$~m and $2,000$~m. The results are shown in
Fig.~\ref{fig:Spacevariation}.

The first observation is that in the free-space case contact and
inter-contact distributions exhibit very close power-laws. This is a
reminiscent characteristic of the isolate behavior~-- individuals
can take the entire place they want. Looking at the runs with
periodic space, the evolution of the distributions suggests a
specific value of density that corresponds to the point of
equilibrium between socialize and isolate behaviors. Above this
point, the results tend to an exponential law, which is an
inheritance of the RWP model. This is probably due to the fact that
individuals move quickly from a point to another, and all the
conditions change dramatically from one simulation step to another,
rendering senseless the decisions of social adaptation. Below the
point of equilibrium, the results obtained with a periodic space
tend to approach the ones of free space.

\subsection{Influence of considered time}

\begin{figure}
  \begin{center}
  \hspace*{-2.4mm}
  \epsfxsize=8.9cm
  \leavevmode\epsfbox{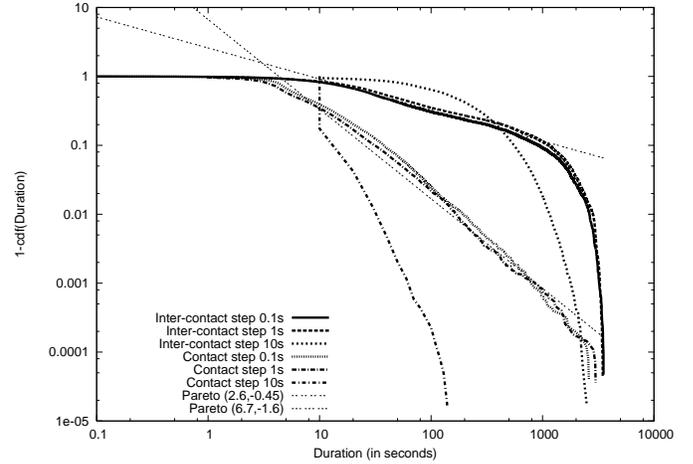}
  \caption{Effect of the time quantization step on SIMPS results.}
  \label{fig:Timequantization}
  \end{center}
\end{figure}

The influence of time on SIMPS outcomes appears in two ways: time
quantization step and total simulation time. The first aspect is
related to the very common problem of sampling on measurements. It
can be summarized in one simple question: ``Is the considered time
step fine enough for a realistic view of the system?'' To explore
this issue, we ran three tests, with times quantization steps of
$0.1$~s, $1$~s, and $10$~s, respectively. The results are shown in
Fig.~\ref{fig:Timequantization}.

It is straightforward to observe that the results with
$\Delta_{\tau} = 0.1$~s and $\Delta_{\tau} = 1$~s are very close one
to the other, while $\Delta_{\tau} = 10$~s completely changed the
results. We can consider then with confidence that a time
discretization step of $1$~s is fine enough for the results we wish
to observe. Such a time granularity is in accordance with the
half-perception time of $\tau_{r} = 4$s we identified in our earlier
experiments, and with the precept that a human goes from
steady-state to full motion or inversely in around $1$~s (recall
that this value was used for acceleration distribution).

\begin{figure}
  \begin{center}
  \hspace*{-2.4mm}
  \epsfxsize=8.9cm
  \leavevmode\epsfbox{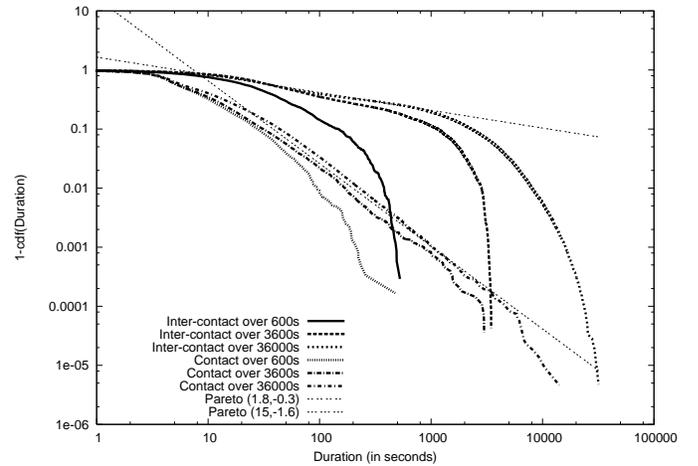}
  \caption{Effect of the total duration considered on SIMPS results.}
  \label{fig:DurationEffectCutoff}
  \end{center}
\end{figure}

The second important aspect related to the influence of time is the
total duration of the simulation, since it directly influences the
cut-off in the tail of the distributions. The fundamental question
here is: ``Do these observations extend with time considered, or are
they limited in scale?'' Again, three tests were conducted, with
total durations of $\tau_{\max}~= 600$~s, $\tau_{\max} = 3,600$~s,
and $\tau_{\max} = 36,000$~s. The results are shown in
Fig.~\ref{fig:DurationEffectCutoff}. As expected, the cut-offs are
shifted right or left depending on the duration of the simulation.
This comforts theoretical results that mobility contact and
inter-contact distributions are purely power-law in essence.

\section{Further discussion}
\label{sec:furthernotes}

\subsection{About complex systems modeling}

The main discussion relative the modeling of complex systems lies in
the differences between two approaches:

\begin{itemize}

\item Simple models with mathematically tractable parameters
should be preferred whenever possible, since their use allows
obtaining strict results and deterministic predictions.

\item More complex models, on the other side, often tend to be
non-deterministic and only render empiric results, on smaller
parameter ranges, with stochastic conditions.

\end{itemize}

The advantages in terms of ``realism'' of the latter approach is
often considered dubious, when simpler models might render similar
predictions. Indeed, a complex model is only interesting when it
explores situations to which deterministic models cannot apply. With
these considerations in mind, we aimed at making SIMPS as simple as
possible, though exhibiting complex emerging features. Although
SIMPS could be made more complex and specialized in many manners
(for example, by introducing a third behavior stating that an
individual just pauses at some location), we believe that such
additions would restrict the scope of our results. Such adaptations,
however, are welcome in more specific situations.

We also believe that SIMPS brings novelty in two ways: (a) it adapts
outer knowledge from sociology that has not been used before and (b)
it generates mobility patterns that spans a wide range of empirical
observations in an extremely robust fashion.

\subsection{About complex behavior emergence}

The combination of very simple interacting rules leads sometimes to
very complex outputs. Such an emergence phenomenon can be observed
for instance via the SIMPS front-end interface (a real-time
graphical interface of the simulator). We can observe for example
situations referred to as {\it bugger's case}: an isolating node
closely pursued by a socializing node. This appears as pairs of
nodes having a highly asymmetric social relation.\footnote{Such
examples can be further observed in recorded videos and traces of
SIMPS motion, freely available on the SIMPS webpage at
\url{http://www-rp.lip6.fr/~borrel/SIMPS}.} Further observations
showed that such situations resolve themselves since they are part
of a more complex collective mechanism.

\section{Conclusion}
\label{sec:conc}

In this paper, we proposed SIMPS, a new mobility model that
addresses the roots of mobility. SIMPS deals with sociological
influences in human crowd motion found in typical situations like
malls, fairs, cafeterias, clubs, beaches, or fora. SIMPS's modular
conception separates motion influences from motion generation, which
allows associating additional mobility influences with little
effort. Based on a human feature thoroughly studied in sociology,
namely sociability, SIMPS defines two behaviors that regulate each
individual's social interaction level.

We evaluated SIMPS using a simple in-range contact model. Under the
light of recent measures of contact and inter-contact duration
distributions, we draw the following surprising conclusions from our
study:

\begin{itemize}

\item Although entirely synthetic, SIMPS's traces show strong
similarities with various heavy-tailed distributions observed in
real-life situations.

\item Such heavy-tailed features are emergent. While SIMPS uses a
social graph for estimating motion, we show that the graph structure
has negligible influence on the results. Decomposing the influence
of the graph into two complementary behaviors, namely socialize and
isolate, we show the crucial role of their interplay in the final
mobility pattern.

\item The investigation of some input parameters seems to bring
clues on the occasional tendency of contact and inter-contact
distributions to show Weibull distribution characteristics.

\end{itemize}

\section*{Acknowledgements}

\addcontentsline{toc}{section}{Acknowledgments}
We would like to thank Mostafa Ammar for his valuable comments that
helped improve this work.

\bibliographystyle{IEEEtran}
\bibliography{borrel}

\begin{thebibliography}{10}
\providecommand{\url}[1]{#1}
\csname url@rmstyle\endcsname
\providecommand{\newblock}{\relax}
\providecommand{\bibinfo}[2]{#2}
\providecommand\BIBentrySTDinterwordspacing{\spaceskip=0pt\relax}
\providecommand\BIBentryALTinterwordstretchfactor{4}
\providecommand\BIBentryALTinterwordspacing{\spaceskip=\fontdimen2\font plus
\BIBentryALTinterwordstretchfactor\fontdimen3\font minus
  \fontdimen4\font\relax}
\providecommand\BIBforeignlanguage[2]{{%
\expandafter\ifx\csname l@#1\endcsname\relax
\typeout{** WARNING: IEEEtran.bst: No hyphenation pattern has been}%
\typeout{** loaded for the language `#1'. Using the pattern for}%
\typeout{** the default language instead.}%
\else
\language=\csname l@#1\endcsname
\fi
#2}}

\bibitem{Perfect05LeBoudec}
J.-Y. {Le Boudec} and M.~Vojnovic, ``Perfect simulation and stationarity of a
  class of mobility models,'' in \emph{IEEE INFOCOM}, Miami, FL, Mar. 2005.

\bibitem{Jardosh05Real}
A.~Jardosh, E.~Belding-Royer, K.~Almeroth, and S.~Suri, ``Real world
  environment models for mobile ad hoc networks,'' \emph{IEEE Journal on
  Selected Areas in Communications}, vol.~23, no.~3, Mar. 2005.

\bibitem{Johnson04Vehicular}
A.~K. Saha and D.~Johnson, ``Modeling mobility for vehicular ad hoc networks,''
  in \emph{ACM Workshop on Vehicular Ad Hoc Networks (VANET)}, Philadelphia,
  PA, July 2004.

\bibitem{Bettstetter01Smooth}
C.~Bettstetter, ``Smooth is better than sharp: a random mobility model for
  simulation of wireless networks,'' in \emph{ACM MSWiM}, Rome, Italy, July
  2001.

\bibitem{Choffnes05Vehicular}
D.~Choffnes and F.~Bustamante, ``An integrated mobility and traffic model for
  vehicular wireless networks,'' in \emph{ACM Workshop on Vehicular Ad Hoc
  Networks (VANET)}, Cologne, Germany, Sept. 2005.

\bibitem{chaintreau.infocom06}
A.~Chaintreau, P.~Hui, J.~Crowcroft, C.~Diot, R.~Gass, and J.~Scott, ``Impact
  of human mobility on the design of opportunistic forwarding algorithms,'' in
  \emph{IEEE Infocom}, Barcelona, Spain, Apr. 2006.

\bibitem{Kotz02Analysis}
D.~Kotz and K.~Essien, ``Analysis of a campus-wide wireless network,'' in
  \emph{ACM Mobicom}, Atlanta, GA, Sept. 2002.

\bibitem{McNett05Access}
M.~McNett and G.~Voelker, ``Access and mobility of wireless {PDA} users,''
  \emph{Mobile Computing and Communications Review}, vol.~9, no.~2, pp. 40--55,
  Apr. 2005.

\bibitem{tuduce.infocom05}
C.~Tuduce and T.~Gross, ``A mobility model based on {WLAN} traces and its
  validation,'' in \emph{IEEE Infocom}, Miami, FL, Mar. 2005.

\bibitem{Musolesi04Social}
M.~Musolesi, S.~Hailes, and C.~Mascolo, ``An ad hoc mobility model founded on
  social network theory,'' in \emph{ACM MSWim}, Venice, Italy, Oct. 2004.

\bibitem{grossglauser.ton02}
M.~Grossglauser and D.~Tse, ``Mobility increases the capacity of ad hoc
  wireless networks,'' \emph{IEEE/ACM Trans. on Networking}, vol.~10, no.~4,
  pp. 477--486, Aug. 2002.

\bibitem{kn:Zhou99GroupSwarmMobility}
B.~Zhou, K.~Xu, and M.~Gerla, ``Group and swarm mobility models for ad hoc
  network scenarios using virtual tracks,'' in \emph{IEEE Milcom}, Monterey,
  CA, Oct. 2004.

\bibitem{hsu.mc2r05}
W.~Hsu, K.~Merchant, H.~Shu, C.~Hsu, and A.~Helmy, ``Weighted waypoint mobility
  model and its impact on ad hoc networks,'' \emph{ACM Mobile Computer
  Communications Review (MC2R)}, vol.~9, no.~1, pp. 59--63, Jan. 2005.

\bibitem{ref:StudyVillage}
M.~Fors{\'e}, ``Les r{\'e}seaux de sociabilit{\'e} dans un village,''
  \emph{Population (French Edition)}, vol.~36, no.~6, pp. 1141--1162, Nov.
  1981.

\bibitem{ref:StudySports}
R.~Laporte, ``Pratiques sportives et sociabilit{\'e},'' \emph{Journal of
  Mathematics and Social Sciences}, vol.~43, no. 170, pp. 74--94, Nov. 2005.

\bibitem{ref:SocialGraphBuildBehavior}
T.~Snijders, ``The statistical evaluation of social network dynamics,''
  \emph{Sociological Methodology}, 2001.

\bibitem{ref:Snijders}
------, \emph{Models for Longitudinal Network Data}.\hskip 1em plus 0.5em minus
  0.4em\relax New-York: P. Carrington, J. Scott, \& S. Wasserman (Eds.),
  Cambridge University Press, 2005.

\bibitem{hong99group}
X.~Hong, M.~Gerla, G.~Pei, and C.~Chiang, ``A group mobility model for ad hoc
  wireless networks,'' in \emph{ACM/IEEE MSWiM}, Seattle, WA, Aug. 1999.

\bibitem{Musolesi06Community}
M.~Musolesi and C.~Mascolo, ``A community based mobility model for ad hoc
  network research,'' in \emph{ACM/SIGMOBILE International Workshop on
  Multi-hop Ad Hoc Networks: from theory to reality (REALMAN)}, Florence,
  Italy, May 2006.

\bibitem{Borrel04Attractors}
V.~Borrel, M.~D. de~Amorim, and S.~Fdida, ``A preferrential attachment
  gathering mobility model,'' \emph{IEEE Communications Letters}, vol.~9,
  no.~10, pp. 900--902, Oct. 2005.

\bibitem{ref:Hall66HiddenDimension}
E.~Hall, \emph{The Hidden Dimension}.\hskip 1em plus 0.5em minus 0.4em\relax
  Garden City, N.Y.: Doubleday, 1966.

\bibitem{Henderson71Crowd}
L.~Henderson, ``The statistics of crowd fluids,'' \emph{Nature}, no. 229, pp.
  381--383, Feb. 1971.

\bibitem{ref:WelfordReactionTimes}
A.~Welford, \emph{Reaction Times}.\hskip 1em plus 0.5em minus 0.4em\relax
  New-York: Academic Press, New York, 1980.

\bibitem{ref:TriggsDrivers}
T.~Triggs and W.~Harris, ``Reaction time of drivers to road stimuli,'' Monash
  University Human Factors Group - Report HFR-12, Tech. Rep., 1982.

\end{thebibliography}

\end{document}